# 面向 GitHub 编程社区的开源许可证选择的分析[*]

吴 欣[1], 武健宇[1], 周明辉[1], 王志强[2], 杨丽蕴[3]

[1](北京大学 计算机科学技术系,高可信软件技术教育部重点实验室,北京 100871)
[2](西南大学 计算机与信息科学学院/软件学院,重庆 北碚 400715)
[3](中国电子技术标准化研究院,北京 100010)
通讯作者:周明辉, E-mail: zhmh@pku.edu.cn

**摘 要**: 开发者通常会为开源软件选择不同的开源许可证来约束其使用条件,以期能有效地保护知识产权和维持软件的长远发展.然而,开源社区现有的许可证种类繁杂,开发者普遍难以了解不同开源许可证间的差异,而现有的开源许可证选择工具要求开发者了解开源许可证相关条款和明确自己的业务需求,这就使得开发者难以做出合适的选择.学术界虽然对开源许可证有广泛研究,但是对开发者选择开源许可证的实际困难并无系统的分析进而缺乏清晰的认知,有鉴于此,本文旨在从开源开发者角度出发,理解他们选择开源许可证的困难,并通过分析开源许可证的组成要素和开源许可证选择的影响因素,来为开发者选择开源许可证提供借鉴参考.本文设计问卷并随机调研了参与 GitHub 开源项目的 200 名开发者.通过对 53 个反馈结果采用主题分析,我们发现,开发者选择开源许可证通常面临条款内容太复杂和考虑因素不确定这两方面的困难.通过分析 GitHub 上 3,346,168 个代码仓库中使用最广泛的 10 种开源许可证,我们建立了包含十个维度的开源许可证框架.借鉴计划行为理论,我们从行为态度、主观规范和知觉行为控制三个方面提出了影响许可证选择的 9 大要素,通过开发者调研验证了它们的相关性,并进一步通过拟合次序回归模型验证了项目特征与许可证选择的关系.我们的研究结果能加深开发者对开源许可证内容的理解,为开发者结合自身需求选择合适的许可证提供决策支持和经验参考,并为实现基于用户需求的开源许可证选择工具提供借鉴.

**关键词**: 开源许可证;开源许可证框架;开源许可证选择;开源许可证选择的影响因素

# Analysis of open source license selection for the GitHub programming community

WU Xin[1], WU Jian-Yu[1], ZHOU Ming-Hui[1], WANG Zhi-Qiang[2], YANG Li-Yun[3]

[1](Department of Computer Science and Technology, Peking University, Beijing 100871, China)
[2](College of Computer Science and Technology, Southwest University, Chongqing 400715, China)
[3](China Electronics Standardization Institute, Beijing 100010, China)

**Abstract**: Developers usually select different open source licenses to restrain the conditions of using open source software, in order to protect intellectual property rights effectively and maintain the long-term development of the software. However, the open source community has a wide variety of licenses available, developers generally find it difficult to understand the differences between different open source license. And existing open source license selection tools require developers to understand the terms of the open source license and identify their business needs, which makes it hard for developers to make the right choice. Although academia has extensive research to the open source license, but there is no systematic analysis on the actual difficulties of the developers to choose the open source license, thus lacking a clear understanding, for this reason, the purpose of this paper is to understand the difficulties faced by open source developers in choosing open source licenses, analyze the components of open source license and the affecting factors of open source license selection, and to provide references for developers to choose open source licenses. In this paper, we conduct a random survey of 200 developers that participated in the open source projects on GitHub through questionnaires. Through Thematic Synthesis of the 53 feedbacks, we found that developers often faced difficulties in the selection of open source license in terms of complexity of terms and unknown considerations. By analyzing the ten open source licenses most widely used in 3,346,168 repositories on GitHub, we created a framework of open source licenses contains 10 dimensions. Drawing on the Theory of Planned Behavior, we put forward 9 factors that affect license selection from three aspects: behavior attitude, subjective norm and perceived behavior control. The relevance of those factors was verified by developer survey. Furthermore, the relationship between project characteristics and license selection is verified by fitting

---

order regression model. The results of our research can deepen developers' understanding of the contents of open source licenses, provide decision support and experience reference for developers to select appropriate licenses based on their own needs, and provide reference for implementing open source license selection tools based on developers' needs.

在过去 30 年里,开源已经成为软件技术创新和软件产业发展的主要模式.与传统开发模式相比,开源开发展现出充分共享、自由协同、无偿贡献、用户创新、持续演化的新特征,颠覆了诸多经典软件工程的基本假设和理论[56],吸引着越来越多的开发者和企业.目前开源软件已经占据很大的市场份额,例如,Netcraft 关于 2020 年 1 月份 Web Server 的市场调研数据[2]显示开源软件 Nginx 和 Apache 分别以 37.70%、23.98%的市场份额位列前二,而 Microsoft 次之,占比为 14.03%.然而,开源软件的开发和使用也伴随着许多的风险,其中最大的风险之一就是潜在的知识产权侵权责任[9].开源许可证规范了软件的使用、修改、重新发布、担保和归属,为保证自己的知识产权被合理利用,开发者通常会为其项目选择合适的开源许可证.

开源许可证的选择对项目的开发和演化及软件的使用和市场至关重要. Linus Torvalds 曾提到开源许可证是 Linux 成功的决定性因素之一,并明确表示对 GPLv2 的偏爱,因为它强制其他开发者进行反馈,有效地防止了碎片化[55].Android 是一个基于 Linux 内核的操作系统,尽管 Linux 内核遵循 GPL,然而 Google 独立开发的用于硬件驱动和 App 接口的中间层 AndroidSDK 却未遵循 GPL 协议,其知识产权被控制在 Google 手中[54],Google 为了吸引更多的厂商参与,使用宽松型许可证 Apache2.0 对 AndroidSDK 进行开源.开发者以及用户非常关心所使用的开源软件的许可证类型,例如,Redis 变更模块开源许可证,从 AGPL 迁移到 Apache2.0 与 Commons Clause 相结合的许可证,对销售许可软件作了限制,在当时引起了极大的争议[58].

目前,开源社区存在大量不同类型的许可证,仅 OSI 认证通过的开源许可证已有 80 多个,尽管每个开源许可证背后的基本原理是相似的,但它们之间存在相当大的异质性[8].一方面,一些开源许可证之间的细微差异常常让开发者感到困惑,例如, Apache2.0 和 Mulan PSL v2[3]都授予专利权,但对于授权主体的范围表述不同,且 Mulan PSL v2 不需要开发者对修改进行声明,从而降低产生法律纠纷的风险也更完善地保护开发者的切身利益.另一方面,不同的开源许可证可能不同程度地吸引志愿者参与软件开发,从而一定程度上影响着开源软件发展,例如,Jorge 等人通过对 SourceForge 托管项目数据的分析发现 copyleft 型开源许可证比非 copyleft 开源许可证吸引了更多的志愿开发者[1]. 而开源许可证的选择是项目负责人将项目开源时首要考虑的配置因素之一,他们通常认为同一种许可证无法满足不同项目的需求[7].

为了帮助开发者选择开源许可证,业界已经实现了一些开源许可证选择工具.ChooseALicense[4]基于几种简单的项目应用场景为开发者提供推荐,例如,对于依赖社区开发的开发者推荐简单且限制少的许可证(如 MIT 开源许可证),而对于偏向于共享源码的开发者建议选择 GPLv3 许可证等,ChooseALicense 的特点是简单易理解,然而其推荐选择的范围单一且局限,也没有对开发者的复杂需求进行考虑.OSSWATCH 的 Licence Differentiator[5]和码云[6]代码

---

[2] https://news.netcraft.com/archives/2020/01/21/january-2020-web-server-survey.html

[3] 北京大学作为国家重点研发计划"云计算和大数据开源社区生态系统"子课题的牵头单位,依托全国信标委云计算标准工作组和中国开源云联盟,联合国内开源生态圈产学研各界优势团队、开源社区以及拥有丰富知识产权相关经验的众多律师,在对现有主流开源协议全面分析的基础上,共同起草、修订并发布了木兰宽松许可证,第 2 版(Mulan PSL v2),并获得了 OSI 认证. http://license.coscl.org.cn/MulanPSL2

[4] https://choosealicense.com/

[5] http://oss-watch.ac.uk/apps/licdiff/

[6] https://gitee.com/

托管平台集成的许可证选择工具,主要原理是基于开源许可证之间的差异,根据开发者对相关条款的偏好进行选择,帮助开发者缩小许可证选择的范围,使用这类工具的前提是开发者必须了解所选择的开源许可证条款的含义.然而,大多数开源许可证中包含大量生涩难懂的法律术语,开发者难以直接从条款内容分析对比,例如,对于很多商业开发者来说准确理解 GPL 的含义通常是一个常见挑战[53].同时,开发者参与开源的实际需求往往各不相同,例如,按照其采用的商业模式获取商业利益、提升其产品的可见性以获取广大用户基础、提高其在开源社区的声誉、获得社区技术支持等,且这些诉求往往不是单一出现的.在与华为等企业的开源专家的访谈中,他们提到希望开源软件在具有良好的兼容性和广大用户的基础上能够避免陷入法律诉讼.而上文提到的 Linux 和 Android 也正是基于不同的诉求而选择了不同的开源许可证.因此,仅仅通过对许可证条款的选择偏好或者基于简单的应用场景的推荐,难以帮助开发者做出最佳决策.

鉴于开源许可证对开源开发的重要影响,软件工程领域对其相关内容已有深入研究,主要体现在对开发者选择或变更开源许可证的动机、以及开源许可证选择对项目成功的影响上,例如,开发者的经济动机决定开源许可证的选择[12],许可证类型影响用户兴趣和贡献者的数量,并与软件开发速度、开发者提交贡献的频率和参与的持久性相关 [1,2]等,一定程度上提高了开发者对开源许可证的认识.然而,随着开源的不断深入,开发者选择开源许可证的困难愈发突出,学术界对开发者选择开源许可证的实际困难并无系统的分析,进而缺乏清晰的认知.因此,本文从理解开发者选择开源许可证所面临的具体困难出发,通过分析许可证组成要素和影响开源许可证选择的因素,来帮助开发者更好选择开源许可证.具体来说,本文主要回答下述三个研究问题:

RQ1:开发者为项目选择开源许可证时通常会面临哪些困难?
RQ2:开源许可证的组成要素有哪些?
RQ3:哪些因素影响开发者选择开源许可证?

本文采用定性和定量相结合的方法回答上述问题,首先通过阅读现有文献及结合有关项目开发经历,我们设计了调查问卷,从 GitHub 项目仓库的作者中选取 200 名开发者进行问卷调研,并基于主题分析的方法总结出开发者选择开源许可证通常遇到的两类困难,分别是①开发者通常难以理解许可证的条款内容,许可证之间的相似性及其复杂的法律含义让开发者感到困惑;②开发者选择开源许可证时通常受到多种因素影响,他们对如何全面考虑各方面因素进行最佳决策感到困惑,例如项目的特征、开源许可证是否被广泛使用以及开源许可证对项目是否产生影响等.本文通过对 GitHub 上使用最广泛的 10 种开源许可证进行对比分析,采用主题分析的方法提取了一个多维度的开源许可证框架,可以帮助开发者认识开源许可证包含的要素和分析开源许可证间的差异.我们借鉴了计划行为理论中的三个维度,通过问卷调研以及相关文献调研,将开发者的业务需求概括为 9 个方面的影响因素,包括:个人的开源理念、对利益因素的评估、所在组织的观念、开源社区对许可证的偏好、其他项目影响、许可证流行度、许可证兼容性、对项目特征及个人能力的评估、以及许可证选择结果的影响等,通过开发者调研验证了它们的相关性,并进一步通过拟合次序回归模型验证了项目特征与许可证选择的关系.最后本文讨论了研究结果的指导意义和应用场景.

本文的主要贡献可以简要总结如下:

①调研并识别了开源许可证选择的两类常见困难:开发者难以理解开源许可证的条款内容且许可证之间的相似性以及复杂的法律含义让开发者感到困惑;开发者选择开源许可证时通常还会综合考虑多方面因素,他们对如何全面考虑各方面因素进行最佳决策感到困惑.②围绕开源许可证核心要素建立了一个开源许可证框架,可以帮助开发者方便理解开源许可证内容的构成以及开源许可证之间存在的差异.③揭示了影响开发者选择开源许可证的 9 大因素,可以指导开发者结合自身业务需求选择合适的许可证.

本文第 1 节对相关工作进行阐述.第 2 节对开源许可证的背景进行介绍.第 3 节探索开发者在选择开源许可证

时通常遇到哪些困难.第 4 节建立开源许可证框架,以帮助开发者认识和了解开源许可证的构成和存在的差异.第 5 节探索影响开发者选择开源许可证的影响因素.第 6 节讨论了研究结果的指导意义和应用场景.第 7 节阐述了本文的局限性.最后进行总结.

## 1 相关工作

随着开源软件越来越广泛地被使用,开源许可证也逐渐受到学术界和工业界更多的关注.目前,关于开源许可证的研究领域主要集中在法律[36],经济管理[50],社会学[11]以及软件工程等相关领域[2];在研究的内容上主要包括开源许可证的选择、开源许可证合规性使用[17,48]以及相关自动化工具,如开源许可证选择工具[47]、开源许可证检测工具[57]、开源许可证管理工具[58]等方面.而其中关于开源许可证的选择的研究主要集中在如下几个方向:

一是开源许可证的选择或变更的影响因素.开源软件开发者的动机一直是与开源软件相关的研究人员和专业人士讨论的主题[23],Halina 和 Mark 认为许可证的选择主要取决于软件开发人员的意图和期望的结果[9];Ravi 等人使用动机和态度理论来研究开发者对三种开源许可证类型的偏好[7];Lerner 和 Tirole 指出开源许可证的选择是由许可方和开发者社区的经济动机所决定的[12];Darren 讨论了不同类型的开源许可证的利益相关者的需求与义务[30];Singh 等人展示了开发者所处的社会环境如何影响他们对开源许可证的选择,以及开发者的个人经历如何调节这种影响[11];Rober 和 Gregorio 研究了开源项目中许可证变更的动机和影响[25]; Vendome 等人通过定量和定性的方法研究了 GitHub 上的 java 项目中许可证何时以及为什么会改变[16].

二是开源许可证的选择或变更对项目成功的影响.Gottfried 等人分析了十年间开源项目的许可证选择和相关项目增长的趋势[27];Stewart 等人调研了开源项目的开发活动以及用户兴趣与许可证类型的关系[2];Jorge Colazo 和 Yulin Fang 基于社会运动理论研究了开源许可证的选择与项目活动之间的关系[1];Medappa 和 Srivastava 着眼于许可证的选择如何影响贡献者的动机并研究了其与项目成功的关系[16];Kashima 等人进行了开源许可证对软件重用影响的定量研究[29];Maria 等人探索了开源许可证之间的依赖关系,用来解释和指导开源项目的许可证选择[17];Chris 和 Walt 通过对 Apache 基金会创建和迁移到 Apache2.0 以及 NetBeans 项目迁移到 Joint Licensing Agreement 过程的案例研究,分析许可证变更所带来的影响[28];Yuhao 等人调研了由于开发者进行了许可证变更而导致包含不同许可证的相同原始文件的不一致性问题[24].

三是许可证选择自动化工具.例如 Kapitsaki 和 Charalambous 等人采用相似用户和相似项目实现了开源许可证推荐[47].此外,还有大量的研究出现在经济法律和管理领域,关注于常见开源许可证的对比分析[8,9,18,19,33,45],Mikko,Stefano 和 Fabio 等人分析了开源公司如何使用双重许可,确定了双重许可的法律和经济要求[36,39].

尽管这些研究提供了一些如何选择开源许可证的见解,但它们通常仅在某一方面或个别细节上进行了深入的分析和建议,还缺乏对开发者实际需求的综合考量和研究,而开源许可证的选择的影响因素可能是多方面的,开发者难以直接将这些研究结论应用于不同的应用场景或变化的业务需求中.同时,业界对开发者选择开源许可证面临的实际困难还缺乏清晰的认识和系统的分析,现有的开源许可证选择工具要求开发者清楚了解开源许可证相关条款的含义和明确自己的业务需求,难以做出合适的推荐.本文工作旨在从开发者角度出发,发掘他们在选择开源许可证过程中遇到的实际困难,提炼出具有一般性和通用性的开源许可证框架以及影响开发者选择的多个角度和因素,为开发者选择开源许可证提供决策支持和经验参考,同时为实现基于用户需求的开源许可证选择工具提供借鉴.

## 2 研究背景

### 2.1 开源许可证的产生

软件的版权保护意味着软件只能在版权所有者的许可下使用[9].通过版权法保护软件是很重要的,因为软件很

容易被复制,但是创造它却是非常困难和代价高昂的[18].目前,版权保护策略主要存在两种形式:私有版权策略和非私有版权策略.

私有版权策略是一种将部分或全部潜在的技术用户排除在外的策略[6].在私有软件开发模型中,代码首先受到版权保护,然后根据授权协议进行分发,从而赋予用户特殊的权利[18].通过授权软件,软件制造商可以限制用户的责任和权利,例如:只允许在一台计算机上使用等[33],而用户需要为使用、分发、复制或编辑软件支付版税.

非私有版权策略主要包括将软件置于公共领域或进行开源许可[6].将软件置于公共领域意味着完全放弃对其软件的版权保护,任何人都可以无偿地使用和修改,甚至可以删除作者的名字视为自己的作品,.软件的私有化被早期一些程序员认为是"不道德"的行为,在 20 世纪 80 年代中期,麻省理工学院的程序员 Richard Stallman 开发了一种新的软件分发方法,即 GNU 公共许可证[12].Stallman 关于自由软件的革命性思想随后演变为当前的开源软件运动,自由/开源软件的主要目的是最大限度地开放以及减少软件使用传播创新的障碍[33].开源许可通过分配知识产权的权利来共享软件代码,用户和开发人员社区可以自由访问,以促进不同动机的参与者之间的合作和有益的交流[32].综上所述,开源软件和私有软件的主要区别在于它们的许可模式.开源软件仍受版权保护,任何想使用或使用它的人都需要许可证.

## 2.2 开源许可证

开源促进会(Open Source Initiative, OSI) [7]创建于 1998 年,是一个审查和批准开源许可证的非盈利组织.他们为开源许可证建立了一套一致的标准,称之为"开源定义"(Open Source Definition, OSD)[12].开源定义对于开源许可证的标准包含以下几个条件:

必须允许任何人以源代码或其他形式重新发布程序,且不收取任何费用;源代码免费可获取,或收取不超过传输成本的费用;必须允许分发衍生或修改后的软件;不歧视任何用户群体或应用领域;保证源代码的完整性;允许对许可证中的权利重新分发;许可证不得特定于产品;许可证不得限制其他软件;许可证必须是技术中立的[8].

通常我们认为符合开源定义的许可证就是开源许可证.此外,OSI 还注册了一个认证标志:OSI 认证标志(The OSI Logo),这个标记可以放在发布的软件上,这样人们就可以很容易地识别出这是开源软件并且所使用的许可证符合开源定义[8].

## 2.3 开源许可证的类型

开源许可证之间最重要的差异,是关于许可证对分发衍生软件的限制性,即当他人对代码修改和扩展(与其他软件合并)后并分发的限制要求.目前,开源许可证按照限制的强弱通常分为三种类型:

- 宽松型(permissive):这类许可证通常只要求被许可方承认原始作者,衍生软件可以成为私有软件,如 BSD 许可证、MIT 许可证、Apache 许可证.
- 限制型(copyleft):旨在促进开发人员的合作,保护源代码的自由共享[43].copyleft 条款要求对软件的修改和扩展,必须按照获得该软件的许可证进行开源,如 GPL 许可证、OSL 许可证.
- 弱限制型(weak-copyleft):弱限制型许可证要求对软件的修改,重新分发必须按照获得该软件的许可证进行开源,然而合并这些软件的大型作品可以成为私有作品.这是一个折中的方法,允许将代码集成到自己的软件中,而不必使整个代码库开源,避免了不得不分享的场景[5].

---

[7] https://opensource.org/

[8] The Open Source Definition, Last modified,2007-03-22, from https://opensource.org/osd.

# 3 开发者选择开源许可证面临的困难(RQ1)

我们在阅读大量的相关文献及与有关企业开发人员访谈后,进一步通过问卷调查(详见附录)的方式,来了解和分析开发者为项目选择开源许可证时通常面临哪些困难.

## 3.1 方法设计

(1) 问卷设计

我们通过阅读大量文献以及网页信息,并咨询华为等企业开发人员后,初步分析了开发者选择开源许可证过程中可能遇到的困难.以此为基础,我们遵循相关性、完整性、互斥性和可能性的原则[57]设计了问题1(Q1)共7个问卷选项(多选,详见表1),且包含开放式回答(other)的选项,给予问卷回复的开发者表达自己的遇到的其他问题或者并没有在开源许可证选择方面遇到困难.选项设计具体理由如下:

首先,很多网站都维护了一个开源许可证列表,如 FSF[9],OSI,SPDX[10]等,开发者获得开源许可证的信息并不难,难点在于开发者如何过滤这些大量信息以获得相关性,以及利用不完全的决策问题来评估这些信息[11].因此我们设计了选项1(*option1*)和选项2(*option2*).另一方面,由于开源许可证的内容缺乏一致性和标准化,例如:MIT 与 GPL 在内容结构上存在相当大的差异,开发者并不总是清楚许可证中授权与限制的含义,而许可证的法律性质加剧了这个问题[46],因此设计了选项3(*option3*).

其次,问卷中选项4(*option4*)的设计是因为在项目的开发过程中,多个项目之间可能存在大量的交互,如链接、合并、代码块复用等.当所依赖的项目使用了不同开源许可证时,开发者不得不考虑许可证的兼容性问题及合规使用问题[17,48].

最后, 选项5(*option5*)和选项6(*option6*)的设计源于已有研究揭示了开源许可证类型与项目的开发活动、用户兴趣的关系[2],以及开源许可证的选择对贡献者的动机和项目的成功影响[16].此外,我们还发现除了 OSI 认证的 80 多个开源许可证外,仍然有大量的不尽相同的开源许可证,中移苏研的开源专家曾提到目前已有超过 2000 个用于开源社区的许可证,出现这种现象的可能原因是现有的许可证无法满足开发者的需求,所以,设计了选项7(*option7*)的选项.

同时,在设计问卷的调查对象时,我们将问题2(Q2)开发者的所在地区(填空)、问题3(Q3)开发者参与开源年限考虑在内(单选),以期望问卷结果具有广泛性.

我们从 GitHub 网站托管的代码仓库中利用 GitHub Search API[11]收集了流行度排名前十的编程语言的项目仓库(创建于 2018 年 1 月至 2019 年 9 月之间)共 9672 个,其中编程语言流行度我们参考了 TIOBE[12]编程语言 2019 年 9 月排行榜,它反映了编程语言流行趋势以及某个编程语言的热门程度.由于这 9672 个项目仓库存在重复的情况,因此我们对数据进一步清洗,通过 ID、创建时间的一致性删去了重复的项目,共得到 7938 个项目仓库.为了便于分析许可证类型,我们又从中剔除了许可证为空和标为 Other(通过人工随机抽查 5 个被标记为 Other 的开源项目,主要包括多重许可证、自定义开源许可证或正在实施许可证变更等情况)的项目仓库,最终得到了本次实验的 4704 个项目.我们从 4704 名项目所有者(Owner)中随机抽取 200 名(我们参考了统计学关于简单随机抽样样本量计算方法[61],按照调查结果在置信度为 95%,误差范围在 4%-8%之间的抽样样本数为 146-533,同时为了避免产生过多打扰,我

---

[9] https://directory.fsf.org/wiki/Category:License

[10] https://spdx.org/licenses/

[11] https://developer.github.com/v3/guides/

[12] TIOBE Index for September 2019, from https://www.tiobe.com/tiobe-index/

们最终确定发放样本 200 份),向他们发送邮件,并附上利用问卷星[13]制作的问卷链接.

(2) 数据分析

我们使用主题分析方法对问卷结果进行分析.主题分析(Thematic Synthesis)是一种识别、分析和报告数据中的模式(主题)的方法,通常用于对定性研究数据进行分类,在软件工程等众多领域广泛使用[40].主题分析一般分为五个步骤①数据提取(Extract data),是指从问卷答复中提取数据;②数据编码(Code data),从数据集中标识和编码感兴趣的概念、类别、发现和结果;③概念化(Translate codes into themes),将标识和编码的内容总结为子主题;④范畴化(Create a model of higher-order themes),探索子主题间的关系并总结为更高阶的主题;⑤验证(Assess the trustworthiness of the synthesis),评估主题分析的解释的可信性.

步骤①中我们通过对收到的回复,提取出选项及补充的其他答案等信息;步骤②我们从前面提取的信息中定位内容的关键点,以系统的方式识别和编码有关开发者选择开源许可证过程中可能遇到的困难(图 1);步骤③中我们通过对识别的开源许可证选择的多个困难对比,总结共性和归纳主题形成概念性的主题;步骤④中我们对前面所归纳的主题再次抽象,将相似问题归为一类,在此基础上总结结论,即开发者选择开源许可证面临的困难;最后是步骤⑤,本文的多个作者对分析的结果进行交叉验证,获得了一致意见.

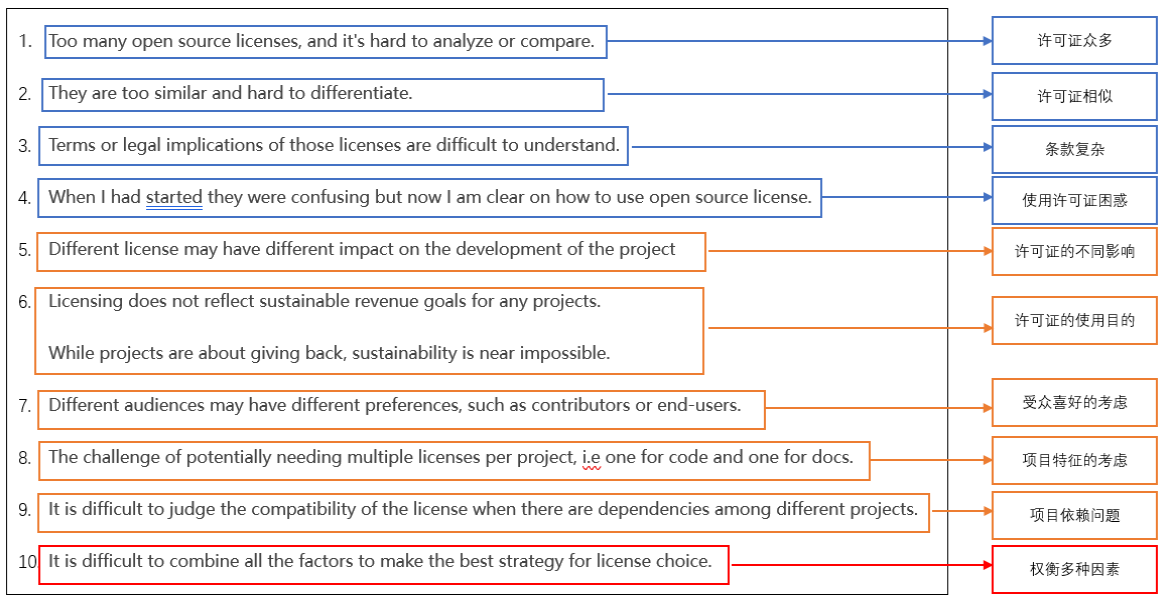

Fig.1　Qualitative analysis of material (Questionnaire options and open reply) encoding process
图 1　定性分析材料(问卷选项及开放式回复)数据编码步骤示例

## 3.2 结果分析

我们向 3.1 中选取的开发人员共发放 200 份问卷,最终收到 53 份回复,回复率为 25%,软工领域中邮件调研回复率通常在 6%-36%[62],因此我们的回复率在相对较高的范围.其中,回复问卷的开源开发者主要来自包括美国,德国,日本,瑞典等 25 个国家,样本的分布广泛.这些开发者中 23%参与开源1-3 年,77%开发者有超过 4 年的开源经验.84%

---

[13] https://www.wjx.cn/

的开发者认为许可证的选择是困难的,其中包含 83.33%的开源经历为 1-3 年的开发者,以及 79.31%的超过 4 年以上开源经历的开发者.仅 8 人(15.38%)表示没有困难,其中 1 人提到 "开始时,开源许可证让我感到困惑,但现在我清楚了如何去使用它们(When I had started they were confusing but now I am clear on how to use open source license)" .这些都进一步说明了开源许可证选择的困难是普遍实际存在的.

我们对得到的 53 份回复结果采用主题分析,最终归纳出两类困难,分别是:

①开发者认为目前存在大量的开源许可证,它们相互之间的相似性使其难以区分,且一些开源许可证的条款和复杂的法律含义使其感到困惑.39.62%的开发者认为目前存在大量的开源许可证,难以对比或分析;而一些开源许可证的条款以及所包含的法律含义让人难以理解,例如 Apache2.0 和 GPLv2 中对专利授权的范围不同,由于 Apache2.0 包含了专利授权终止条款,使得 Apache2.0 无法兼容 GPLv2.一名开发者在补充回答中提到"它们太相似了,很难区分(They are too similar and hard to differentiate)".开源许可证是相似的,它们之间可能存在一些共同的模式,找到这种模式可能帮助开发者更好理解和对比开源许可证;

②开发者选择开源许可证时可能综合考虑多种因素,例如,他们可能针对不同的项目或者项目特征为其选择适用的开源许可证,开发者还关心许可证选择的结果是否影响项目的发展等等,开发者对如何全面考虑各方面因素进行最佳决策感到困惑.33.96%的开发者认为许可证选择的结果可能进一步影响项目的发展,例如 MIT、BSD-2-clause、Apache2.0 等开源许可证在分发义务中允许衍生软件成为商业软件,可能使得软件具有更多的用户,这种影响的不确定性增加了他们选择开源许可证的难度;28.3%的开发者在选择开源许可证时容易受到许可证兼容性问题的困扰,例如,在 Linux kernel 基础上扩展软件需要遵从 GPLv2 的许可证约束,而判断所选的开源许可证是否与其兼容需要一定的法律知识以及了解许可证的内容;16.98%的开发者认为,针对不同的项目受众应当选择不同的许可证,例如一些具有自由开源理念的贡献者通常会选择具有限制型许可(如 GPL 类许可证)的项目进行贡献,一名来自美国,超过 10 年的开源开发经验的人认为"每个项目可能需要多个许可证,比如代码部分使用一个许可证,文档部分使用另一个许可证(The challenge of potentially needing multiple licenses per project, i.e. one for code and one for docs)",也进一步说明了开发者选择开源许可式会考虑针对项目的不同类型选择不同的许可证;还有 15.09%的开发者认为影响开源许可证选择的因素是多方面的,如何全面考虑这些因素进行选择是困难的.

此外还有开发者提到"对项目进行许可并不实现可持续发展的目标,虽然开源项目是为了回报社会,但其可持续性几乎是不可能的(Licensing does not reflect sustainable revenue goals for any projects. While projects are about giving back, sustainability is near impossible)" 表达了开源许可证具有一定的局限性,尽管其规范了开源软件的使用、复制修改和分发,但无法保证项目的可持续发展.值得注意的是,option7 结果为 0,出现这个结果原因可能是,我们选取的调研对象是从去掉许可证为空或其他许可证的项目作者中挑选,而所调研的开发者认为目前的开源许可证已经能够满足他们的需求.

**Table 1** The difficulties developers face in choosing an open source license

表 1 开发者选择开源许可证面临的困难

**(问题 3)在选择开源许可时遇到了什么困难?(多项选择题)**

**(Q3)What difficulties have you met in choosing an open source license? (multiple-choice)**

| 选项 | 数量 | 比例 |
| --- | --- | --- |
| 1.开源许可证种类太多,难以进行分析和比较(Too many open source licenses, and it's hard to analyze or compare). | 21 | 39.62% |
| 2.难以综合所有的因素来做出最佳的许可证选择策略(It is difficult to combine all the factors to make the best strategy for license choice). | 8 | 15.09% |
| 3.许可证中的术语或法律含义很难理解(Terms or legal implications of those licenses are difficult to understand). | 21 | 39.62% |
| 4.当不同项目之间存在依赖关系时,很难判断许可证之间的兼容性(It is difficult to judge the compatibility of the license when there are dependencies among different projects). | 15 | 28.30% |
| 5.不同的许可证可能会对项目的发展带来不同的影响(Different license may have different impact on the development of the project). | 18 | 33.96% |
| 6.不同的受众对许可证可能有不同的偏好,比如贡献者或最终用户(Different audiences may have different preferences,such as contributors or end-users). | 9 | 16.98% |
| 7.现有的许可证无法满足自己的需求(Exising licenses cannot meet our needs). | 0 | 0.00% |
| 8.其他(Other). | 11 | 20.75% |

### 3.3 结论

我们发现①大部分的开发者在为项目选择许可证是困难的,尤其是新参与的开发者在面对众多开源许可证,它们内容的相似性和复杂的法律含义让人感到困惑.②开发者选择开源许可证时通常还会综合考虑多方面因素,他们对如何全面考虑各方面因素进行最佳决策感到困惑.

## 4 开源许可证的组成要素(RQ2)

为了帮助开发者更好地理解开源许可证,以及减少开发者选择许可证的困难,我们通过主题分析的方法探究开源许可证的组成要素.

### 4.1 方法设计

首先,我们使用谷歌 Bigquery[14]工具获取了目前开源许可证的使用情况.谷歌 Bigquery 是一种用于处理和分析大数据的 Web 服务,其提供的 GitHub 的公共数据是迄今为止最大的 GitHub 可用数据源,所收录的项目仓库均受到一个开源许可证约束.截止至 2019 年 11 月 25 日,我们分析了共 3,347,168 万个项目仓库, 提取每个项目授权的许可证信息,并统计每种开源许可证的使用情况.

其次,我们选取广泛使用的前十种许可证,采用主题分析[40]的方法进行分析:其中,数据提取步骤中我们对选取的十种许可证,提取出许可证的基本信息,条款内容以及使用说明等;数据编码的步骤中,我们从前面提取的信息中

---

[14] https://console.cloud.google.com/

定位内容的关键点,以系统的方式识别和编码许可证的内容(图 2);概念化的步骤里,通过对许可证内容的关键信息对比,总结共性和归纳主题形成概念性的主题,将许可证内容划分为多个维度;范畴化的步骤里,我们对前面所归纳的多个维度再次抽象,总结出开源许可证框架.

最后,为了检验本文提出的维度是否能反映大部分开源许可证的内容,本文作者通过人工分析 OSI 或 FSF 认证的 72 个许可证内容对提出的维度进行交叉验证.

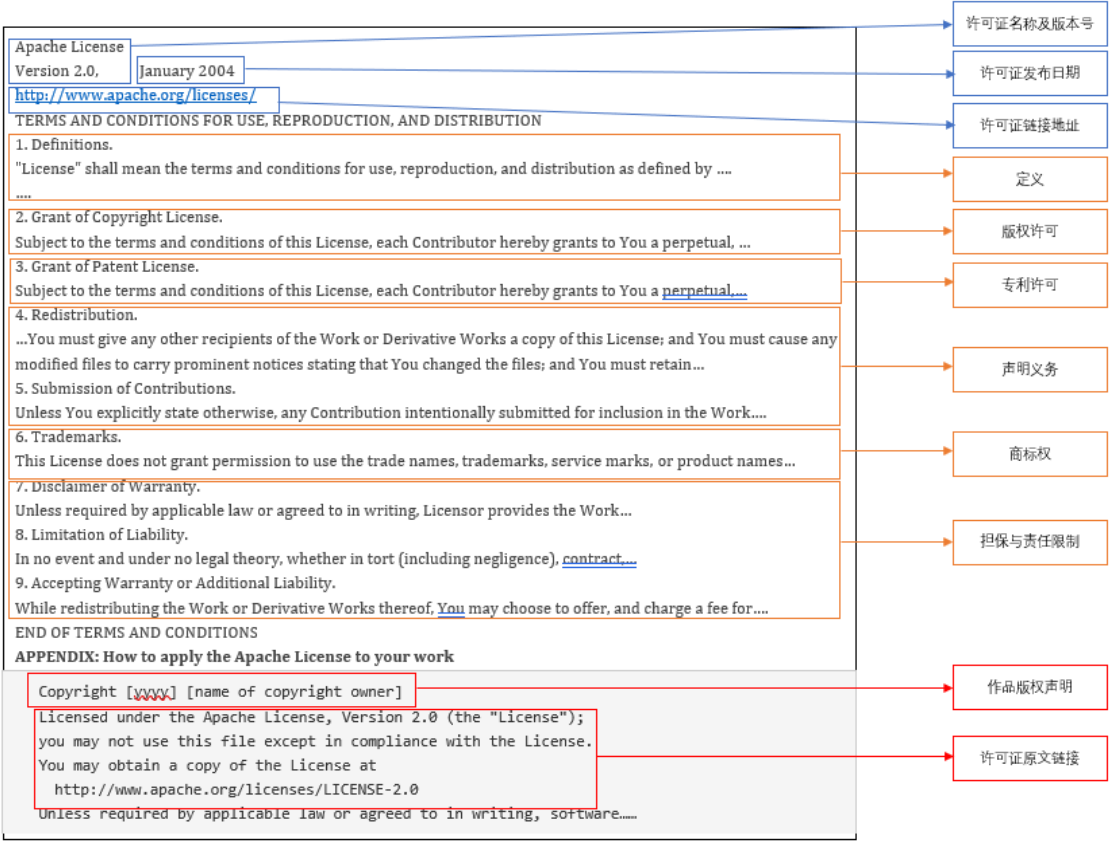

Fig.2  Qualitative analysis of material(part of Apache2.0) encoding process
图 2  定性分析材料(截取部分 Apache2.0)数据编码步骤示例

4.2  结果分析

我们对 3,346,168 个许可证信息进行统计,得到图 2 的结果.由图中可知 MIT,Apache-2.0,GPL-3.0,GPL-2.0,BSD-3-Clause,BSD-2-Clause,AGPL-3.0,LGPL-3.0,CC0-1.0,EPL-1.0 这十种开源许可证是使用最广泛的,占比 97%.其中 MIT 许可证使用最多,占比 51%;其次 Apache-2.0 占比 15%;GPL-3.0 和 GPL2.0 各占比 10%.开源社区中普遍使用的开源许可证主要是 OSI 对许可证分类中的"流行且广泛使用的许可证".按照开源许可证的类型进行划分,其中 MIT,Apache-2.0,BSD-3-Clause,BSD-2-Clause,CC0-1.0,ISC,Artistic-2.0 属于宽松型许可证,LGPL-3.0,EPL-1.0 属于弱限制型许可证,而 GPL-2.0,GPL-3.0,AGPL-3.0 属于限制型许可证,可以看出宽松型许可证占主导地位(总计 76%),限制型许可证次之(21%),而弱限制型许可证并不常见(4%).

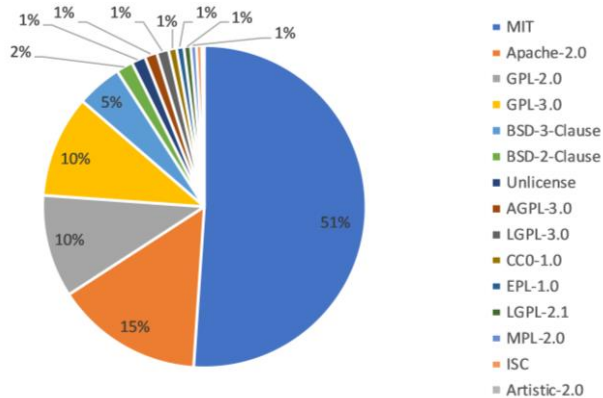

Fig.3　Open source license usage on GitHub
图　3　GitHub 上开源许可证使用现状

随后我们对选取的十种许可证的内容进行主题分析,将许可证条款划分为 10 个维度:①许可证的基本信息,②序言,③定义条款,④授权条款,⑤义务条款,⑥违约与授权终止条款,⑦担保与责任限制条款,⑧准据法条款,⑨许可证版本与兼容性,以及⑩许可证使用说明等.其中①许可证的基本信息、④授权条款中的版权许可、⑤义务条款、⑦担保与责任限制条款等为开源许可证中的常见条款,其他条款通常根据许可证制定方的需求进行相应说明.具体如下:

1) 基本信息. 主要包括许可证名称及版本号、发布日期、许可证版权声明及链接地址等信息.
2) 序言. 主要对许可证的适用场景或条件以及目的宗旨等进行说明,如 GPL 序言部分.
3) 定义. 为了便于开发者或用户理解许可证内容,对许可证条款中的特定术语进行说明.
4) 授权. 开源许可证主要涉及的知识产权主要包括版权、专利权、以及商标权.
   - 版权许可. 项目开发者通常免费授予用户行使相关权限,包括使用、复制、修改、分发其开源项目或修改后的项目.
   - 专利许可. 如果开源项目中包含专利, 开发者可以提供专利许可, 也可以不提供专利许可.对于不提供专利许可的场景,许可证将不会提及专利许可,或者明确排除专利许可.
   - 商标权. 开源许可证原则上不涉及商标权许可,并且通常禁止借项目开发者的名义进行广告或宣传.
5) 义务. 用户在使用、复制、修改或分发软件或其衍生软件应当遵守的行为规范.
   - 使用/复制/修改. 开源许可证通常不对使用目的、范围进行限制,开发者可以基于任何目的(学习、研究或商业)对软件进行运行、备份、修改等操作,但在一些特定场景中可能要求履行相关义务.如 AGPL 中对通过网络使用软件向第三方提供服务时,需要提供完整源代码.
   - 分发. 当开发者分发软件或衍生软件时通常要求,①履行一定的声明义务,例如修改声明(如 Apache2.0)、保留版权及免责等声明、提供许可证副本等;②要满足开源许可证对分发衍生软件的限制性,根据限制性的强弱,可以将开源许可证分为三类,即宽松型(permissive)、弱限制型(weak-copyleft)、限制型(copyleft).
6) 违约与授权终止. 对于用户违反许可证的行为,可以对其终止授权,也可以给予一定的补救机会.
7) 担保与责任限制. 项目开发者通常不对用户提供任何担保以及承担任何赔偿责任;如果开发者个人对用

户提供保证和担保须自行承担相应责任.
8) 准据法. 是指在许可证中指定援用的,用来调整涉外民事法律关系双方当事人权利与义务的特定国家的法律.
9) 版本与兼容性. 对许可证版本进行说明,以及对与该许可证兼容或者不兼容的其他许可证进行特别说明,许可证兼容性是指项目中的许可证包含相互矛盾的必要条件,而使得无法将其源代码合并成新的项目.
10) 使用说明. 告知用户将项目许可在该许可证下应该完成的步骤,并可以包含一个声明模板,对软件相关的作品描述、作品版权声明、许可证及其链接、作者联系方式等进行说明.

开源许可证之间的差异通常体现在③定义、④授权、⑤义务、⑥违约与授权终止以及⑧准据法中,例如:MPL2.0 和 CPAL1.0 对"覆盖代码"定义的范围不同,因此在下文条款中要求对"覆盖代码"使用原许可证分发其源码,MPL2.0 可以通过不同文件来隔离传染性,CPAL1.0 则需要不同模块单独分发来隔离传染性;在授权中 Apache2.0 明确提供授予专利权,而 MIT 则未提及专利授权;在义务条款中常见的差异主要来源于对分发衍生软件的限制性强弱以及不同的声明义务;在授权终止中一些开源许可证中可能包含专利报复条款以及不同的违约补救条件,如 Apache 中提到用户不得发起专利诉讼,否则其授权将被终止;此外,还有一些开源许可证明确了准据法,如 EUPL1.2,这也是开发者在使用开源许可证过程中需要注意的问题.

最后,我们根据这十个纬度,提出了开源许可证框架(图 4),本文的作者通过对开源许可证进行交叉验证,认为提取的开源许可证框架可以较好地解释目前开源许可证的结构组成,能够帮助开发者理解许可证的条款.

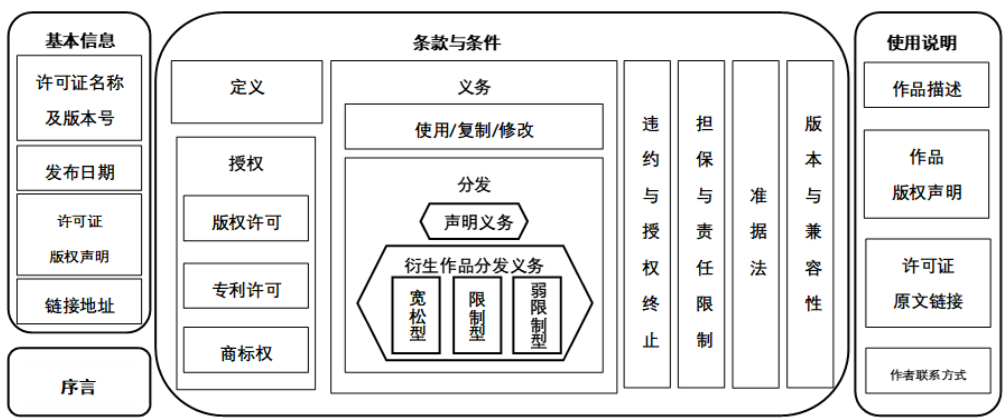

Fig.4　Open source license framework
图 4　开源许可证框架

### 4.3　结论

我们发现,目前在 GitHub 上广泛被使用的是 OSI 分类为"流行且广泛使用"的开源许可证,其中宽松型开源许可证占主导地位,限制型次之,而弱限制型的开源许可证并不常见.通过对开源许可证内容进行主题分析,提出了十个维度的开源许可证框架,其中①许可证的基本信息、④授权条款中的版权许可、⑤义务条款、⑦担保与责任限制条款等为开源许可证中的常见条款,其他条款通常根据许可证制定方的需求进行相应说明.开发者可以通过比较各个维度上开源许可证之间的差异,来理解开源许可证的结构与含义.此外,开发者还可以利用开源许可证框架快速构建满足个人特定需求的新开源许可证,例如,木兰宽松许可证,第 2 版(Mulan PSL v2)正是在此开源许可证框架基础上为更好地保护开发者权益而构建的中国本土开源许可证.

## 5 开源许可证选择的影响因素(RQ3)

为了剖析影响开发者选择开源许可证的因素,进一步为开发者结合自身需求选择合适的许可证提供决策支持和经验参考,我们通过阅读有关文献并借鉴计划行为理论(Theory of Planned Behavior)的三个维度设计调查问卷,分析开源许可证选择的影响因素,通过对项目特征因素拟合次序回归模型,验证分析了项目特征与开源许可证选择的关系.

### 5.1 方法设计

(1) 问卷调研

在设计 2.1 中面向 200 开发者的调研问卷时,我们同时调研了开发者选择开源许可证的考虑因素.在设计问卷问题过程中,我们借鉴计划行为理论的三个维度,结合有关文献,初步分析了可能影响开发者选择开源许可证的因素. 计划行为理论是从信息加工的角度、以期望价值理论为出发点解释个体行为一般决策过程的理论,大量研究证实它能显著提高研究对行为的解释力和预测力[14].在许可证选择过程中,我们认为所有可能影响许可证选择的因素都是经由开发者的行为意向来间接影响最终许可证的选择.这点经常被现有文献验证,例如程序员的参与意愿强烈影响他在社区中的持续性[44,59].行为意向是指个人想要采取某一特定行为的行动倾向,受到三项相关因素的影响:一是个人本身的"行为态度";二是外在的"主观规范";三是"知觉行为控制"[37].以此为基础,我们遵循问卷设计中相关性、完整性、互斥性和可能性的原则[57]设计了问题 4(Q4)、问题 5(Q5)、问题 6(Q6)、问题 7(Q7)、问题 8(Q8)、问题 9(Q9)、问题 10(Q10).其中,问题 4(Q4)调研开发者选择开源许可证的影响因素,共有 9 个选项(多选,详见表 2),并增加一个 other(开放式回答)选项,收集开发者关于许可证选择因素的补充回答;问题 5(Q5,开放式)调研开发者是否为项目变更过许可证及变更原因;问题 6(Q6,单选)调研开发者具体支持何种开源理念;问题 7(Q7,单选)调研开发者的首选开源许可证;问题 8(Q8,单选)调研开发者如何看待开源商业;问题 9(Q9,多选)调研开发者采用的商业模式;问题 10(Q10,多选)调研开发者通常获得哪些开源利益.为了保证问卷设计的合理性及计划行为理论维度的适用性,在发送调查问卷前,我们与 10 名有丰富开发经验的科研及企业人员进行了调研讨论,详细地解释了研究目的和调研问卷设计的方法思路,并明确要求他们填写问卷以检查问卷中可能包含的问题.我们根据收集的建议对问卷中问题及选项的描述进行了澄清和改进.实际问卷调研的 53 份回复不包含前述的实验性预调研结果.

问卷及(Q4)选项设计的理由如下:

①行为态度方面,Ajzen 等人提出的期望-价值理论[38]认为态度包括个人实行某种行为的重要信念以及对价值的评价.自由软件哲学思想的传播在自由软件和 GPL 的普及过程中发挥了重要作用[19],开发者通常选择最符合自身意愿的开源许可证,例如,支持限制型许可证的人们认为共享代码是一个理想的最佳实践,可以允许程序员创建更高质量的软件[20],因此我们设计了选项 1(*option1*).近年来,人们对开源软件作为一种替代经济模式的兴趣日益浓厚,一些公司通常会寻找新的方式来产生收入和降低成本,越来越多的公司将开源作为一种商业策略来实现这个目标[34],受此影响,我们设计了选项 2(*option2*).

②主观规范方面,是指个体在决策时感知的社会压力或者外界因素的影响.当开发者不是以个人名义参与开源时,通常可能受到来自组织或企业的影响,例如,开发者在 Linux 社区中使用 BSD 许可证是可以接受的,但是他们在 BSD 社区中贡献 GPL 代码是一个大大的禁忌[5],由此我们设计了选项 3(*option3*).贡献者和用户对许可证类型的不同偏好可能影响开发者选择开源许可证,例如,具有限制性许可证的项目吸引的主要是寻求高度内在动机的贡献者,而具有宽松许可证的项目吸引的不仅是寻求内在动机的贡献者,还有希望获得商业化潜在机会的贡献者[16],根据这一点我们设计了选项 4(*option4*).许可证的流行度和复杂度也可能影响开发者是否采用该许可证,一个众所周知的和受信任的许可证对于开发者和用户来说都更容易被接受,而过于复杂和鲜为人知的许可证可能容易造成混淆和

歧义[17],所以,我们设计了选项 5(*option5*)和选项 6(*option6*).此外,项目之间的依赖使得许可证的选择可能会受到其他许可证的限制,当开发者试图将他人编写的代码集成到自己的项目中时,他们需要了解集成代码所携带的许可证,尤其是组合不同许可证下的代码时,兼容性问题会变得格外复杂[33],有鉴于此我们设计了选项 7(*option7*).先前的研究发现,项目采用的许可证类型往往受到与其关系密切的其他项目采用的许可证类型的影响[6],由此我们设计了选项 8(*option8*).

③知觉行为控制方面,是指个人预期采取某一特定的行为时所感觉可以控制的程度,常反映个人过去的经验、拥有的资源、能力以及预期的阻碍等.从 3.2 节中我们得知,开发者在为项目选择许可证时会考虑不同的受众目标,且有文献表明开发复杂软件的项目更可能选择有某种程度限制的许可证[7],开发者可能依据不同的项目特征选择不同的开源许可证,例如,Oracle 公司在 GitHub 上托管的不同开源项目中采用了多种开源许可证,如微服务框架 Helidon 使用 Apache2.0 开源,而另一款移动交友软件 DinoDate 则使用 MIT 许可证开源,有鉴于此我们设计了选项 9(*option9*).此外,已有研究表明开源许可证的选择与项目的开发活动有关[2],一定程度影响项目的发展.开发者可以通过判断项目的发展趋势是否符合自己的意愿,或者为了使许可证的选择满足自己变化的需求,采用更换开源许可证以达到其目的,因此我们设计了一道开放式问题 5(Q5,开放式)调研开发者是否为项目更换过许可证和更换原因.同时,为进一步了解开发者的具体偏好,我们针对上述选项 1(*option1*) 设计了问题 6(Q6,单选,表 3):关于开发者具体支持何种开源理念的问题,及问题 7(Q7):开发者首选的开源许可证;针对上述选项 2(*option2*)设计了问题 8(Q8,单选,表 4):关于开发者如何看待开源商业,问题 9(Q9,多选,表 5):开发者采取了哪些商业模式,以及问题 10(Q10,多选,表 6):开发者从开源中通常获得哪些开源利益等问题.

(2) 定量分析

大量的工作探究了开源许可证类型与项目之间的关系[1,2,16],也有研究利用相似用户或相似项目为开发者推荐开源许可证[47],项目特征可能是影响开发者选择开源许可证的重要因素之一.为了验证项目特征与许可证选择有关,我们通过拟合次序回归模型分析了项目特征与许可证类型之间的关系,通过数据分析及有关文献调研解释了可能原因,为开发者提供参考借鉴.具体步骤如下:

已有研究提出了几个重要的项目特征,包括项目年龄、目标受众、编程语言、项目大小等[2],我们对 2.1 中选取的 4704 个项目仓库,提取出项目的创建日期、编程语言、应用程序描述、项目大小、许可类型、以及项目的开发数据等有关信息.我们共设置了四类自变量,分别为编程语言、应用领域、项目规模和项目年龄.其中:①编程语言(*PL*):包括 *C, C#, C++, Java, JS, Objective-C, PHP, Python, SQL, VB*;②应用领域(*Domain*):通过人工分析项目的应用程序描述, 我们根据不同的受众,将项目应用领域分为 6 类:软件开发类(*Develop*,包括开发工具、库/框架等)、终端应用类(*App*,包括桌面应用、web 应用、移动应用等)、流行技术类(*Popular*,包括人工智能、云计算、数据科学、区块链等)、底层相关类(*Underlying*,包括操作系统、数据库、中间件、硬件相关等)、教程(*Tutorial*)及游戏(*Game*).③项目规模(*Size*):我们利用 GitHub 上项目的存储大小信息按照分布的百分位数分为 3 个层次:其中前 1/3 为小项目(<=1M),中间 1/3 为中项目(1M-20M),后 1/3 为大项目(>20M);④项目年龄(*Age*):以月为计量单位,按照创建时间的先后进行统计.因变量为许可证类型(*LicenseType*=1,2,3),分为宽松型、弱限制型以及限制型.自变量中的编程语言和应用领域为无序分类变量,使用两组虚拟变量表示,其中编程语言中的 *VB* 和应用领域中的游戏(*Game*)为参照类,我们将上述变量带入次序回归模型进行拟合.次序回归模型定义如下式:

$$LicenseType \sim PL + Domain + Size + Age$$

## 5.2 结果分析

根据计划行为理论的三个维度结合有关文献调研,我们提出影响开源许可证选择有 9 个因素,其中①开发者的

开源理念、②开发者对利益的评估等涉及行为态度方面;③开发者所在组织观念的影响、④社区偏好的影响、⑤许可证流行度和复杂度、⑥许可证兼容性、⑦其他项目的影响等涉及主观规范方面;⑧对项目特征的评估、⑨许可证选择结果的影响等涉及知觉行为控制方面.问卷调研结果验证了各个影响因素的相关性.我们进一步通过拟合次序回归模型验证了项目特征与开源许可证选择的关系,具体分析如下:

Table 2 Factors affecting developers' choice of open source licenses

表 2　开发者选择开源许可证的影响因素

| (问题 4)在选择开源许可证时,您主要考虑的是什么?(多项选择题) | | |
|---|---|---|
| (Q4) What are your main considerations when choosing an open source license? (multiple-choice) | | |
| 选项 | 数量 | 比例 |
| 1.取决于个人的意愿(According to your own will). | 25 | 47.17% |
| 2.考虑所采用的商业模式(Consider the business model used). | 16 | 30.19% |
| 3.所在组织的指导原则会影响许可证的选择(The guiding principles of your organization influence decision). | 7 | 13.21% |
| 4.考虑社区中开发者或用户的偏好(Preferences of developers or users from Community). | 17 | 32.08% |
| 5.考虑许可证的流行度,是否被广泛使用(License popularity, whether such license is widely used). | 13 | 24.53% |
| 6.考虑许可证的法律复杂度(Legal complexity of the license). | 12 | 22.64% |
| 7.许可证之间的依赖,依赖包的许可类型影响许可证的选择(License dependency, license choice may be limited to the license type of dependency packages). | 9 | 16.98% |
| 8.相似项目的许可证选择方案(Existing license choices for the similar projects). | 7 | 13.21% |
| 9.考虑项目的特征,如采用的编程语言、应用领域、项目规模等(Characteristics of your project (such as programming language, application domain, project size)). | 9 | 16.98% |
| 10.其他(Other). | 5 | 9.43% |

①开发者的开源理念.

开源理念反映了开发者如何看待开源软件的使用问题,当开发者认为某一个开源许可证所蕴含的开源哲学与其开源理念相符时,更有可能表现出强烈的选择意愿.问题 4(Q4)选项 1(*option1*)反映了开发者的开源理念(表 2),调研中发现 47.17%的开发者认为个人意愿是影响他们选择开源许可证的一个因素.此外开发者还在补充回答中提到"自由/开源哲学理念(FLOSS philosophy)"、"为了防止其他人将我们的项目用于商业产品,最大化耗费开源资源(To prevent others taking our project and using it in a commercial product, maximize consumption)".也进一步说明了开源理念在许可证选择过程中的重要影响.

关于如何开放源码,不同的开发者有不同的理念,在问题 6(Q6)的调研(表 3)中,我们得到在开发者支持何种开源理念的调研结果中,35.85%的开发者支持知识共享观念,一定程度上说明他们中许多人不希望源代码被第三方私有化,而是期望获得最大的贡献和反馈[9],可能更愿意使用限制型许可证;24.53%的开发者则支持更多最终用户权利的开发人员更喜欢限制较少的开源许可证[7];而 39.62%的开发者没有明显的倾向性,他们更有可能受到其他一些因素的影响.

**Table 3** Developers' philosophy about open source

表 3 开发者的开源理念

| (问题 6)你最认同哪一种观点? | | |
|---|---|---|
| (Q6) Which viewpoint do you agree with most? | | |
| 选项 | 数量 | 比例 |
| 1.软件应该对所有用户自由的,它可以被每个人共享和修改(Software should be free to all users, it could be shared and modified by everyone). | 19 | 35.85% |
| 2.用户有更多的权利以及更少的限制(Users might have more rights and fewer restrictions). | 13 | 24.53% |
| 3.介于两者之间(In-between). | 21 | 39.62% |

在问题 7(Q7)的关于开发者首选开源许可证的调研结果(图 4)中,显示 75.47%的开发者选择了宽松型开源许可证,25.53%的开发者选择了限制型开源许可证,而没有开发者选择弱限制型开源许可证,总体上与 GitHub 数据集观察到的许可证使用情况的分布相吻合.我们通过交叉图分析开发者支持的开源理念与其首选开源许可证,发现支持知识共享的开发者相比于支持更多用户权利的开发者更容易选择限制型的开源许可证.然而,开发者的开源理念并不完全决定其选择的开源许可证类型,一部分支持知识共享观念的开发者选择了宽松型开源许可证,而另一部分支持更多用户权利的开发者也选择了限制型的开源许可证,说明开发者选择开源许可证时通常受到除开源理念以外其他方面因素的影响.

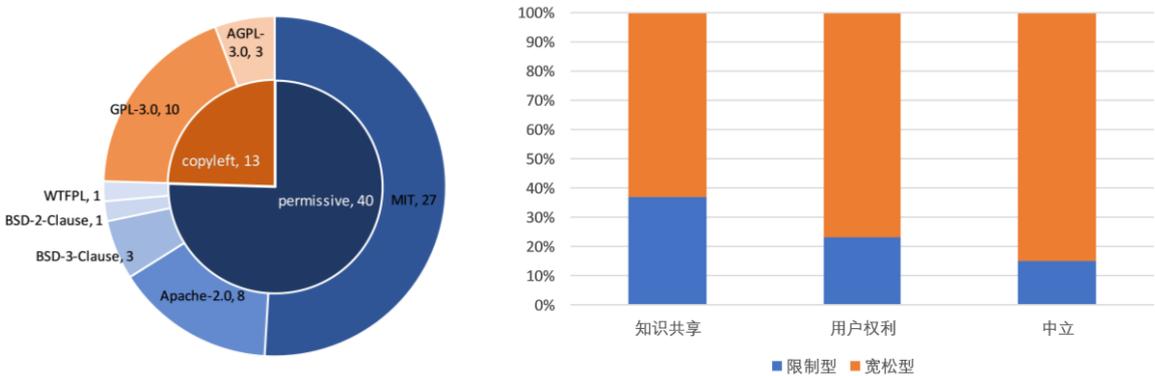

Fig.5 Preferred open source license for developers

图 5 开发者首选的开源许可证

②开发者对利益的评估.

开源并不一定出于利他主义或意识形态,它也可以出于健康的自身利益[6].开发者参与开源获得的利益可以分为经济利益以及非经济利益.

一是经济利益方面,虽然开源软件第一眼看上去似乎是反利润的,但越来越多的企业或组织已经将开源思想纳入到他们的商业战略[52].(表 2)问题 4 选项 2(*option2*)的结果显示 30.19%的开发者认为其采用的商业模式是选择开源许可证的考虑因素.在问题 8(Q8)关于开发者如何看待开源商业的调研结果中(表 4)发现,69.81%开发者支持开源商业,认为可以促进开源的发展,也有 13.21%开发者反对通过开源获得经济利益,认为这是不道德的,而 16.98%的

开发者保持中立态度.

**Table 4** Developers' view on open source commercialization

表 4 开发者如何看待开源商业化

| 选项 | 数量 | 比例 |
|---|---|---|
| **(问题 8)您如何看待开源软件的商业化?换句话说,就是通过使用开源软件来获得经济利益.** <br> **(Q8) What do you think about the commercialization of open source software? In other words, to gain economic benefit by using open source software.** | | |
| 1.支持.它可以促进开源的发展(Support.It could promote open source development). | 37 | 69.81% |
| 2.反对.这是不道德的(Oppose.It would be immoral). | 7 | 13.21% |
| 3.介于两者之间(In-between). | 9 | 16.98% |

针对支持开源商业和保持中立态度的 46 名开发者,我们继续调研他们是否采用以及采用了何种商业模式(表5),调研发现,其中 69.57%的开发者在实际开发中采用了商业模式,包括:双重授权、支持服务、互补产品、闭源销售和赢在声誉等.应用最广泛的是生产互补产品(43.48%),是指基于开源的计算或服务项目提供商业的插件或配套硬件等,如,销售 CD 版的 Linux,销售 Eclipse 的商业插件[35].其次是支持服务(32.61%),是指针对开源软件项目为客户提供支持、维护、开发、咨询或培训服务,或提供与开源相关的审计和法律服务等,例如 mLab 为用户提供数据库项目的托管服务,RedHat 为付费用户提供"知识产权保障计划"[35].再次是双重授权(30.43%),通常包含两种方式:一是在开源许可下提供有限的或精简版的软件产品,而在专有许可证下提供增强的或升级版的软件产品[35],开源免费版主要用于扩大产品可见性和知名度,例如开发工具 Pycharm 同时提供社区免费版以及增强功能的企业版;二是针对同一个项目同时提供开源许可和商业许可,这里采用的开源许可证通常具有 copyleft 特性,主要是为了扩大用户群及降低竞争对手生产盗版私有产品的机会,而为那些希望不受 copyleft 影响的用户提供商业许可,例如 MySQL 和 Sleepcat 的商业策略.而后是赢在声誉(26.09%),为了推广技术或在行业中建立新的标准,将其项目开源,不仅有利于后续产品的推广,还可以从商标、广告中获取经济利益.最后是闭源销售(6.52%),是指将开源软件与私有软件相结合,并作为私有软件销售,如 AWS 提供了 Apache Hadoop 软件的商业版本,然而仅靠销售软件获利并不容易,开源社区已经提供了该产品的免费版本,只有能够为产品增加相当大的价值时,才能产生可观的利润[34].

Table 5 Business model used by developer

表 5  开发者采用的商业模式

**(问题 9)你采用的开源业务模式是?**
**(Q9) Your open source business model？(multiple-choice)**

| 选项 | 数量 | 比例 |
| --- | --- | --- |
| 1.无(None). | 14 | 30.43% |
| 2.双重授权,提供定制服务(Dual authorization, providing customization service). | 14 | 30.43% |
| 3.提供技术培训或售后服务(Provide technical training or after-sales service). | 15 | 32.61% |
| 4.生产基于开源软件的互补产品(Complementary products based on open source software). | 20 | 43.48% |
| 5.闭源销售(Closed-source commercialization, close source code and sell software). | 3 | 6.52% |
| 6.获得声誉(Win the reputation). | 12 | 26.09% |
| 7.其他(Other). | 1 | 2.17% |

二是非经济利益方面,开发者参与开源还可能有经济利益以外的目的,开发者可以通过选择不同的许可证类型以期获得不同的非经济利益,例如限制型许可证可以提供更高的贡献可见性,开发者更容易获得期望的认可、声誉或职业机会[18],而宽松型许可证对用户没有限制,容易获得更大的用户群 [5].问题 10(Q10)调研发现(表 6),开发者通过开源获得的非经济利益主要包括提升个人名誉(81.13%)、获得挑战乐趣(71.7%)、社区开发者的技术支持(58.49%)、职业发展(52.83%)、提升产品可见度(58.49%)、建立行业标准(32.08%)、广泛的用户基础(41.51%)、还有开发者提到"我还用开源库发表了论文(I also published the libraries as papers)"、"企业间合作的投资安全(Safety of investment for collaboration between companies)".

Table 6 The non-economic benefits of open source

表 6  开源的非经济利益

**(问题 10)开源给你带来什么好处?(多项选择题)**
**(Q10) What benefits does open source bring to you? (multiple-choice)**

| 选项 | 数量 | 比例 |
| --- | --- | --- |
| 1.个人声望(Personal reputation). | 43 | 81.13% |
| 2.产品可见性(Product visibility). | 31 | 58.49% |
| 3.建立行业标准(Establishment of industry standards). | 17 | 32.08% |
| 4.挑战的乐趣(Challenge fun). | 38 | 71.7% |
| 5.职业发展(Career development). | 28 | 52.83% |
| 6.扩大用户基础(Broad user base). | 22 | 41.51% |
| 7.社区中开发者支持(Support of developers from community). | 31 | 58.49% |
| 8.其他(Other). | 2 | 3.77% |

③开发者所在组织观念的影响.

当开发者代表组织或在特定社区进行开源时,通常也需要考虑所选择的许可证是否符合其所在组织的观念,而避免不必要的纠纷,例如上文中提到的在 BSD 社区中应避免选用 GPL 类开源许可证.表 2 选项 3(*option3*)的结果

显示 13.21%的开发者认为其所在组织的观念可以影响其选择开源许可证.

④社区偏好的影响.

不同的开源许可证类型不同程度地吸引开源社区中的贡献者和用户,例如:宽松型许可证允许与私有软件合并,因此对商业用户更有吸引力,而一些社会性质项目的开发者可以通过选择限制型许可证来吸引更多的开发人员[7],因为限制型许可证往往产生更强的社会认同感[11];Sen 等人还发现,受到工作挑战激励的人更喜欢有适度限制的许可证,而那些重视诸如地位或机会之类外在动机的人更喜欢宽松的许可证.(表 2)问题 4 选项 4(*option4*) 结果表明 32.08%的开发者认为贡献者和用户对不同类型许可证的偏好影响其选择开源许可证,

⑤许可证流行度和复杂度.

(表 2)问题 4 选项 5(*option5*)结果表明 24.53%的开发者慎重考虑了开源许可证是否被广泛使用,开发者通常被建议使用现有的经过实践检验的许可证,而不是起草一个新许可证.而选项 6(*option6*) 结果表明 22.64%开发者认为许可证法律复杂度是影响其选择开源许可证的重要因素,例如开源许可证中准据法的条款使得开发者不得不考虑其知识产权被合理使用的范围.

⑥许可证兼容性.

(表 2)问题 4 选项 7(*option7*) 结果表明 16.98%的开发者认为项目之间的依赖导致的许可证兼容性问题是其选择开源许可证时考虑的因素,这种情况常常发生在使用了限制型开源许可证的项目.开发者在补充回答中也明确指出"许可证的权限,如项目的副本应以原始项目为核心,采用相同的许可证(Restrictions and permission of the license (copy of the project should core the original project, stays under the same license))" .

⑦其他项目的影响.

(表 2)问题 4 选项 8(*option8*) 结果表明 13.21%的开发者为项目选择开源许可证时可能考虑与其类似项目或者是其他大型知名开源项目所使用的开源许可证.例如,开发者补充说到"我只是模仿大型库的许可方式,比如 scikit learn(I just mimic what big libraries are doing, e.g. scikit learn)"等.

⑧对项目特征的评估.

(表 2)问题 4 选项 9(*option9*)的调研结果发现仅 16.98%的开发者会根据项目的特征考虑不同的开源许可证,一定程度说明项目特征因素在开发者实际选择开源许可证并不重要.

鉴于项目特征对开发者的重要性和可评估性,我们分析了项目特征与开源许可证类型的具体关系,结合文献调研解释其原因,为开发者参考项目特征选择开源许可证提供参考.我们通过对 3.1 中选取的 4704 个项目进行定量分析.我们从项目中提取编程语言、应用程序描述、大小(单位为 kb)、创建时间和许可证信息,并根据 5.1(2)中对变量设定原则得到编程语言(PL)、应用领域(Domain)、项目年龄(Age)和许可证类型(LicenseType)等信息(数据示例如表 7 所示),并进行相关关系分析(表 8)及次序回归拟合(表 9).回归模型通过了平行线检验,说明模型是有效的,项目特征与许可证类型有关,伪 R 方表示项目的四个特征对许可证类型的解释程度, $R^2$ (Cox and Snell)值为 0.169,说明开源许可证的选择还受到项目的其他特征或者项目特征以外的其他因素的影响,与上述调研结果是相符的,也一定程度地说明了仅仅通过相似项目为开发者推荐许可证是不全面的.

**Table 7** An example of the dataset

表 7　数据集示例

| 项目名称 | 应用程序描述 | 大小 | 创建时间 | 许可证 | 编程语言 | 应用领域 | 项目年龄 | 许可证类型 |
|---|---|---|---|---|---|---|---|---|
| Programming-nu/nu | Nu is an interpreted Lisp that builds on the Objective-C runtime and Foundation framework. | 4337 | 2008/03/05 | Apache-2.0 | Objective-C | 软件开发 | 141 | 宽松型 |
| Phusion/passenger | A fast and robust web server and application server for Ruby, Python and Node.js. | 50170 | 2008/03/27 | MIT | C++ | 软件开发 | 140 | 宽松型 |
| Steveicarus/iverilog | Icarus Verilog. | 23305 | 2008/05/12 | LGPL-2.1 | C++ | 硬件相关 | 139 | 弱限制型 |
| … | … | … | … | … | … | … | … | … |

　　从表 7 的相关系数和表 8 的回归模型数据中可以发现,项目年龄与许可证类型的关系并不显著,可以认为在近一年里许可证类型的使用情况变化不大.项目规模与许可证类型的关系是正向显著的,表明项目规模越大,越容易选择限制较强的许可证,正如 FSF 的常见问答中也建议对于代码量较小的项目可以使用限制较少的许可证,而且项目规模越大,开发者可能投入了更多的努力,限制型许可证可以更好地保护开发者的努力不被第三方获取.

　　从编程语言和应用领域这两个分类变量的置信区间估计值及回归系数(表8)可以观察到,编程语言C(0.04)、C++(-0.31)、SQL(-0.34)、VB(参考类)相比于其他语言C#(-1.37)、Java(-1.92)、JS(-2.64)、Objective-C(-2.18)、PHP(-0.97)、Python(-0.87)更容易选择限制型许可证.正如我们所了解的,C、C++通常用于服务端的服务程序开发、硬件和系统开发等,而SQL用于数据库领域,VB主要用于游戏开发以及对软件进行二次开发,相比于Java、Python、C#等倾向于应用开发的语言,前者更偏向于面向底层开发或者其面向特定的受众.从软件应用领域来看,软件开发类(-1.16)、流行技术类(-0.99)、教程(-0.95)等领域更容易选择宽松型许可证,而在终端应用类(-0.47)、底层相关类(-0.29)、游戏(参考类)等领域限制型许可证比在前述其他领域中的限制型许可证更常见.交叉分析统计图(图6)可以直观对比不同类型的开源许可证分别在编程语言和应用领域上的差异,可能的原因是,一方面由于终端用户很少需要源代码,而对于大众市场的软件开发人员通常把核心产品的源代码作为公司盈利的宝石,使用限制性较强的许可证可以避免竞争对手从访问源码中获得搭便车的好处[20],另一方面对于底层相关领域的技术实现相对复杂,开发者通常不希望源码被第三方私有化,且采用限制型许可证可以避免因克隆产生过多分支版本而影响原版本的发展和权威.

**Table 8** Correlation coefficient

表 8 相关系数

|  | **M** | **SD** | 1 | 2 | 3 | 4 | 5 |
|---|---|---|---|---|---|---|---|
| 1. *Age* | 58.36 | 27.830 | 1 | -0.067** | -0.209** | 0.032* | -0.007 |
| 2. *PL* | 5.18 | 2.648 | - | 1 | 0.032* | -0.246* | -0.004 |
| 3. *Domain* | 1.91 | 1.341 | - | - | 1 | 0.092** | 0.229** |
| 4. *Size* | 2.02 | 0.730 | - | - | - | 1 | 0.060** |
| 5. *LicenseType* | 1.32 | 0.720 | - | - | - | - | 1 |

注意:列标题中 1,2,3,4,5 分别表示项目年龄、编程语言、应用领域、项目规模和许可证类型. *p<0.05.**p<0.01.***p<0.001

**Table 9** Estimate of parameter

表 9 参数估计值

| | | *LicenseType* | | 95%CI | |
|---|---|---|---|---|---|
| | | coefficient | wald | upper | lower |
| *Age* | | 0.006 | 13.425 | 0.003 | 0.009 |
| *Size* | | 0.320*** | 27.498 | 0.201 | 0.440 |
| *PL* | *C* | 0.040 | 0.068 | -0.260 | 0.339 |
| | *C#* | -1.365*** | 62.205 | -1.704 | -1.026 |
| | *C++* | -0.308 | 3.661 | -0.624 | 0.008 |
| | *Java* | -1.921*** | 88.197 | -2.322 | -1.520 |
| | *JS* | -2.635*** | 115.166 | -3.117 | -2.154 |
| | *Objective-C* | -2.180*** | 108.830 | -2.590 | -1.771 |
| | *PHP* | -0.966*** | 28.856 | -1.318 | -0.613 |
| | *Python* | -0.867*** | 26.086 | -1.200 | -0.534 |
| | *SQL* | -0.344 | 1.408 | -0.913 | 0.224 |
| | *VB* | 0a | - | - | - |
| *Domain* | *Develop* | -1.159*** | 34.861 | -1.543 | -0.774 |
| | *App* | -0.472* | 6.051 | 0.096 | 0.849 |
| | *Popular* | -0.998*** | 17.500 | -1.466 | -0.531 |
| | *Underlying* | -0.295 | 1.800 | -0.726 | 0.136 |
| | *Tutorial* | -0.954** | 10.150 | -1.541 | -0.367 |
| | *Game* | 0a | - | - | - |
| df | | 16*** | | | |
| N | | 4704 | | | |
| $R^2$ | Cox and Snell | 0.169 | | | |
| | Nagelkerke | 0.258 | | | |
| | McFadden | 0.174 | | | |

*p<0.05.**p<0.01.***p<0.001

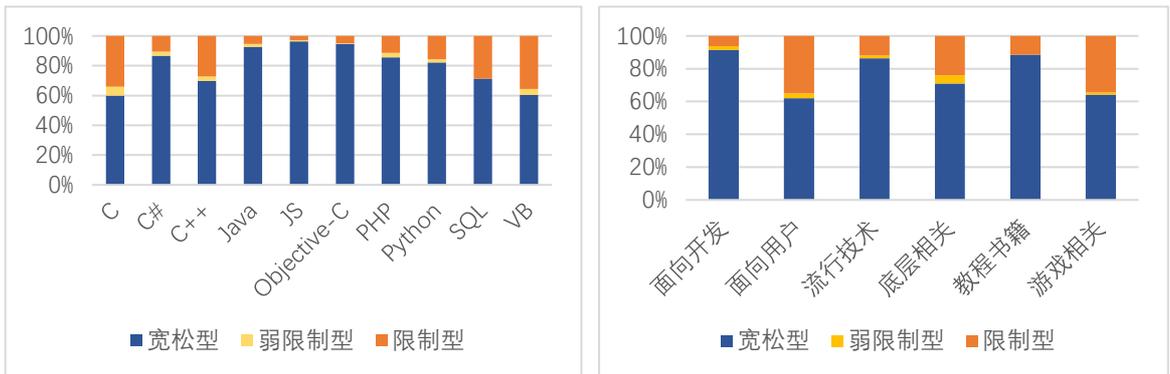

Fig.6　Distribution of open source license type usage in programming languages and domains

图 6　开源许可证不同类型在编程语言和应用领域的使用分布

⑨许可证选择结果的影响.

由于社区的贡献者和用户对开源许可证类型存在不同偏好,一定程度影响着他们是否参与或接受该项目,且不同开源许可证的类型可以带来不同的开源利益,从而对项目的发展带来一定的影响.一方面,开发者可以通过分析其他项目许可证选择结果的影响,为自己选择开源许可证提供决策支持;另一方面,开发者可以根据个人开源项目发展趋势或经验进行判断,调整对开源许可证的选择.我们在问题 5(Q5)调研了他们是否为项目变更开源许可证及变更原因,53 名回复者中有 7 名开发者(13.20%)回复了该问题,表明他们为项目更换过许可证,主要原因包括支持项目的商业化、使项目更容易被接受以扩大用户群、通过更强的限制以达到保护项目不被商业化的目的,以及尽量避免项目因许可证涉及法律纠纷等("I have changed from BSD license to Apache license as the former one is more suitable for EU laws", "I have modified some projects, mainly "Libraries", from GPL-v3 to LGPL-3.0 to ease project acceptance.", "Switched from GPL 2.0 after learning about its complexities and issues", "We had things in gpl that we moved to MIT as that is useable in commercial projects", "I changed from GPL to AGPL to protect my application's API's as well.", "To broaden the consumer audience", "Switched from closed source to open 5 years ago.").而前文中提到的关于 Redis 变更自研模块的许可证, Redis Labs 的联合创始人兼首席技术官 Yiftach Shoolman 表示变更的目的是为了保护开发者的利益和开源软件的持续发展."多年来,云提供商通过销售基于开源项目的云服务,可从中获利数亿美元,可这些项目实际上并不是他们自己开发的,如 Docker,Elasticsearch,Hadoop,Redis 和 Spark.这阻碍了社区投资开发开源代码,因为任何潜在的好处都归云提供商而不是代码开发人员或他们的赞助商[58]".

## 5.3　结论

我们发现,开发者的开源理念、对利益因素的评估、开发者所在组织的观念、开源社区对许可证的偏好、许可证流行度和复杂度、许可证兼容性、其他项目的影响、开发者对项目特征的评估,以及许可证选择结果的影响,都可能在某种情况下影响开发者为项目选择许可证(图5),且这些因素不是单一影响开发者选择开源许可证的,开发者通常受到多方面因素的影响.首先行为态度方面是影响开发者选择开源许可证过程中最常见的影响因素,其中支持 copyleft 观念的开发者相比支持更多用户权利的开发者更可能选择限制型开源许可证,同时开源不仅可以为开发者带来声誉、认可等非经济利益,还可以带来经济利益,越来越多的企业和个人已经将商业模式应用到开源中,包括生产互补产品、支持服务、双重授权、赢在声誉、闭源销售等,开发者可以根据其对不同利益类型的偏好及所采用的不同商业模式考虑不同的开源许可证;其次,尽管开源许可证的选择是属于个人行为,但开发者所处社会环境的

影响也是其选择开源许可证重点考虑的方面,例如所在组织的观念、社区偏好及开源许可证因素的影响;最后,知觉行为控制方面,大量研究关注于开源许可证与项目之间的关系,然而我们通过调研和定量分析发现,项目特征一定程度影响开源许可证的选择,但其并不是开发者实际选择开源许可证时必然考虑的因素,开源许可证选择工具需要从开发者实际需求出发,通过综合判断分析,才能为开发者推荐适用的开源许可证.

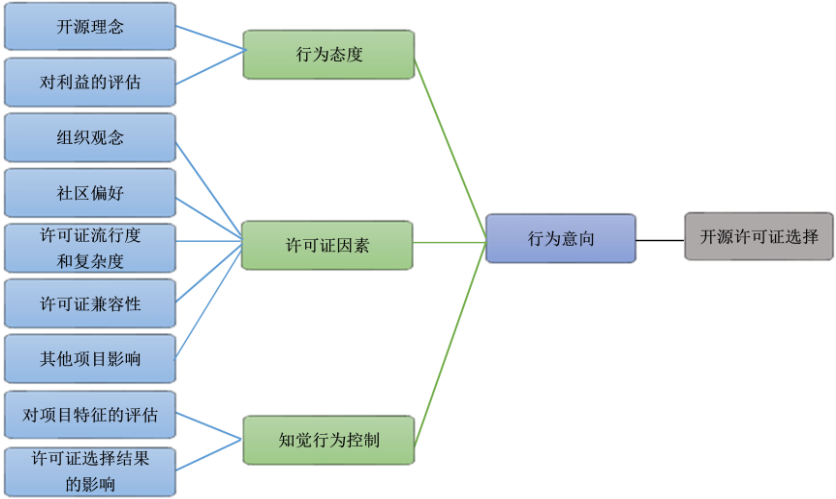

Fig.7　Factors influencing the choice of open source license

图 7　开源许可证选择的影响因素

## 6　讨论

本文针对开发者难以根据自身需要选择合适的许可证的现状,研究了以下问题:①开发者为项目选择开源许可证时通常会面临哪些困难?②开源许可证的组成要素有哪些?③哪些因素影响开发者选择开源许可证?我们通过问卷调研 200 名 GitHub 开源开发者,获得了开发者选择开源许可证通常面临的两类困难和 9 大影响因素,并通过分析 GitHub 开源项目中使用最广泛的 10 种开源许可证,建立了一个开源许可证框架,我们还针对开发者普遍关心的项目特征这一影响因素,通过对项目特征因素拟合次序回归模型,验证了项目特征与开源许可证选择的关系.

我们建立的开源许可证框架可以帮助开发者理解开源许可证的构成和差异.新参与的个体开源开发者通常缺乏法律知识背景,开源小企业可能没有专业的法务团队支持,开源许可证之间的差异以及条款中包含的法律词汇常常使他们感到困惑,我们建立的开源许可证框架可以帮助开发者基于 10 个条款维度分析开源许可证,更容易理解不同开源许可证之间的差异,例如在专利授权条款中有无对违反专利授权的限制可以体现出该开源许可证对开发者利益保护程度;根据在分发义务条款中对衍生作品分发的限制可以将开源许可证划分为宽松型、弱限制型和限制型.开源开发者或开源企业还可以利用开源许可证框架快速构建满足自身需求的开源许可证.尽管开源社区已有大量不同类型的开源许可证,这些开源许可证仍然可能无法满足开源开发商的所有需求,他们试图构建自己的开源许可证,例如,MongoDB 改用一种新的服务器端公共许可证(SSPL)力求堵住一些云提供商利用其开源代码生产数据库的托管商业版本而无需开源的缺口[15].因此,我们提出的开源许可证框架还可以帮助开发者、社区或开源企业快

---

速构建满足自身需求的开源许可证,在此框架基础上构建的木兰宽松许可证,第 2 版(Mulan PSL v2)通过 OSI 认证也进一步证明了该框架的有效性和通用性.

我们揭示的影响开源许可证选择的 9 大因素可以帮助开发者或开源企业从不同角度全面分析自身的需求来指导开源许可证的选择.开源企业参与开源的目的与其需求直接相关,不同开源许可证的类型可以带来不同的开源利益,且一定程度影响贡献者和用户是否参与或接受该项目,从而对项目的发展带来一定的影响,企业对自身业务需求的准确把握对开源许可证的选择至关重要,因此,开源许可证的选择要综合考虑各方面因素.例如,Raúl Kripalani 在分析了 35 家企业 75 个开源项目的许可证后得出唯一使用相同许可模式的公司是 Palantir(硅谷一家数据挖掘公司)[60].这反映出开发者通常依据自身不同需求针对不同的开源项目在不同的应用场景下选择不同的开源许可证.本文结合问卷和文献调研得到的影响开发者选择开源许可证的9个因素反映了开发者选择开源许可证不同方面的考虑,我们建议开发者可以从 9 个影响因素的维度全面准确分析自身需求,在充分理解开源许可证条款的基础上做出综合判断,从而选择符合自身需求的开源许可证.

此外,我们的研究结果还可以帮助业界清晰了解开发者选择开源许可证面临的实际困难和影响因素,采用更好的策略来解决开发者遇到的困难,例如改进现有的开源许可证选择工具来帮助开发者结合自身需求选择合适的许可证提供决策支持.目前业界和学术界实现的基于开源许可证条款差异(OSSWATCH Licence Differentiator)、基于简单应用场景(ChooseALicense)、或基于相似项目推荐[47]的开源许可证选择工具,没有全面考虑到开发者的实际困难和需求,难以帮助开发者选择适用的开源许可证.我们结合定量分析和调研也发现尽管项目特征和开源许可证类型存在一定的相关性,但其并不是开发者选择开源许可证时必然考虑的因素,开源许可证选择工具需要从开发者实际需求出发,通过综合判断分析,才能为开发者推荐适用的开源许可证.本文得到的 9 个影响因素维度为实现基于用户需求的开源许可证选择工具提供了思路,通过识别开发者的需求,借鉴相同或相似需求的成功或著名开源项目所选择的开源许可证,为开发者推荐适用的开源许可证.我们的未来研究方向是如何获取和利用开发者的不同需求,结合不同开源项目的特征和应用场景,为开发者推荐合适的开源许可证.

## 7 局限性

论文的局限性主要体现在数据的有效性、方法的构造性和结论的普适性上面.

第一个局限性是数据的可访问性和一致性.首先,本文通过 GitHub search API 爬取的项目数据只包含有限项目仓库,我们获取了编程语言流行度排名前十的总共 9672 个项目仓库,这些项目仓库包含了近一年的数据,一些仓库没有公开或者无法通过 GitHub search API 获得,在数据清洗过程中,我们去掉了许可证为空或 Other 的项目,因其无法提取项目的许可证信息;其次,数据本身可能无法反映实际的情况.在 5.2 中我们定量分析了项目特征与许可证类型之间的关系,其中项目的特征仅考虑了编程语言、应用领域、项目规模和项目年龄,可能存在其他的特征影响着许可证的选择,例如项目的技术复杂性等.尽管这可能导致结果分析存在一定的局限性,但编程语言的流行度与软件开发领域发展和市场热度息息相关,通常是开发者关注的热点领域,而对最近一年数据的开源许可证使用情况分析也正好反映开发者选择开源许可证的现状.因此,我们认为所获取的数据能够适用于本文研究工作的目的.

第二个局限性是方法的构造性.首先,本文主要采用主题分析方法和计划行为理论等定性分析方法对调研结果进行分析,然而采用定性研究结果指导实践具有一定的局限性;其次,在分析开源许可证选择影响因素时,我们认为行为意向能较好地解释开发者选择开源许可证的行为,因此借鉴了计划行为理论的三个维度设计调研问卷,然而从计划行为理论的研究进展可以发现,计划行为理论的主要变量的概念定义一直是研究者们争论的焦点[14].这可能对分析结果的准确性造成一定的影响.我们通过预调研的方式确保问卷设计的合理性及计划行为理论维度的适用性,并对定性分析的结果进行交叉验证从而减少这类影响.

第三个局限性是结论的普适性.在调研中随机选取的 200 名开发者作为调研对象不能代表所有开发者的意见,由于邮件调研的特殊原因,一部分开发者的邮箱地址为公司客服邮箱或者已注销或不再使用,难以得到有效的回复,且开发者容易受到当时环境、心理等因素的影响,可能存在其他未调研到的实际困难和影响因素,使研究结果的应用范围和应用程度受到一定的局限.由于国际邮件调查的回复率普遍较低,在软件工程领域,邮件回复率通常在 6%-36%[62],而我们的回复率在 25%,是一个相对较高的回复率.同时,我们通过阅读大量文献、与相关企业开发人员访谈设计问卷、预调研等多种方式确保问卷的合理性和选项设计涵盖的普遍性,通过实际问卷调研来获取更广泛的反馈并细致分析结果,最终得到开发者选择开源许可证时面临的困难和影响因素.我们认为所得研究结果可以反映当前开发者普遍面临的困难及其选择开源许可证过程中主要的考虑因素,具有普适性.

## 8 总结

本文中我们首先通过问卷的方式调研了开发者在为项目选择许可证时通常遇到两类困难,包括因许可证的相似性和法律复杂性造成开发者难以理解开源许可证之间的差异、开发者对如何全面考虑各方面因素进行最佳决策感到困惑,帮助人们清晰和全面理解开发者选择开源许可证面临的实际困难.我们通过对比分析最广泛使用的十种开源许可证的条款,将许可证内容划分为 10 个维度,并建立了开源许可证框架,可以帮助开发者清晰认识和理解开源许可证的内容构成,便于对各个维度的对比分析开源许可证之间的差异,对帮助开发者解决第一类困难起到促进作用.同时,开发者还可以利用开源许可证框架快速构建符合自己特定需求的新开源许可证.最后,我们通过对开发者选择开源许可证考虑因素的调研和分析,得出影响开源许可证选择存在多方面因素,包括开发者的开源理念、对利益的评估、组织观念和社区偏好的影响、许可证流行度或兼容性问题、其他项目的影响、个人对项目的评估以及许可证选择结果对项目的影响,为开发者结合自身需求选择合适的许可证提供决策支持和经验参考,可以较好地帮助开发者解决第二类困难,为实现基于用户需求的许可证选择工具提供借鉴.

**附录：关于开源许可证选择的调查问卷**

Dear open source developer,

I am a researcher from Peking University. My colleagues and I are investigating how developers choose open source license for their projects. We expect our study could help to understand what factors affect the license choice and how they affect, and to provide deeper insight into license choice. I would highly appreciate your response that may take you a few minutes.

From GitHub repositories data sets, we know that you are an outstanding developer and have opened your projects. Here are several questions towards license choice seeking your advice:

**Q1. What difficulties have you met in choosing an open source license for your projects?**

A. Too many open source licenses, and it's hard to analyze or compare.

B. Terms or legal implications of those licenses are difficult to understand.

C. Different license may have different impact on the development of the project.

D. Different audiences may have different preferences, such as contributors or end-users.

E. Existing licenses cannot meet our needs

F. It is difficult to judge the compatibility of the license when there are dependencies between different projects.

G. It is difficult to combine all the factors to make the best strategy for license choice.

H. Other____________________.

**Q2. How long have you participated in open source projects?**

A. 1-3years.

B. 4-10years.

C. Over 10years.

**Q3. Your country or region?____________.**

**Q4. What are your main considerations when choosing an open source license for your project? (multiple choice)**

A. According to your own will.

B. Consider the business model used.

C. The guiding principles of your organization influence decision.

D. Preferences of developers or users from Community.

E. Characteristics of your project (such as programming language, application domain, project size).

F. Existing license choices for the similar projects.

G. License popularity, whether such license is widely used.

H. Legal complexity of the license.

I. License dependency, license choice may be limited to the license type of dependency packages.

J. Other____________________.

**Q5. Have you changed the project license and why?**

**Q6. Which viewpoint do you agree with most?**

A. Software should be free to all users, it could be shared and modified by everyone.

B. Users might have more rights and fewer restrictions.

C. In-between.

**Q7. What is your preferred open source license for a project?**

A. Apache-2.0

B. AGPL-3.0

C. BSD-2-Clause

D. BSD-3-Clause

E. CC0-1.0

F. CDDL-1.0

G. EPL-1.0

H. GPL-2.0

I. GPL-3.0

J. MIT

K. MPL-2.0

L. Other______________________.

**Q8. What do you think about the commercialization of open source software？In other words, to gain economic benefit by use open source software.**

A. Support. It could promote open source development.

B. Oppose. It would be immoral.

C. In-between.

**Q9. Your open source business model?**

A. None.

B. Dual authorization, providing customization service.

C. Provide technical training or after-sales service.

D. Complementary products based on open source software.

E. Closed-source commercialization, close source code and sell software.

F. Win the reputation.

G. Other______________________.

**Q10. What benefits does open source bring to you?**

A. Personal reputation.

B. Product visibility.

C. Establishment of industry standards.

D. Challenge fun.

E. Career development.

F. Broad user base.

G. Support of developers from community.

H. Other______________________.

**Q11. Do you have any additional comments on open source license choice?**

**Your answer is extremely important for our study! Please note your information will be kept privacy and we will only publish common statistics.**
**If you have any further questions or concerns, please do not hesitate to contact me.**
**Thanks a lot for offering any insights! Sorry if this bothers you.**